\documentclass[%
reprint,
superscriptaddress,
 amsmath,
 amssymb,
 aps,
 amsfonts,
prl
]{revtex4-2}

\usepackage{graphicx}% Include figure files
\usepackage{dcolumn}% Align table columns on decimal point
\usepackage{bm}% bold math
\usepackage{physics}
\usepackage{natbib}
\usepackage{mathtools,slashed}
\usepackage{hyperref}% add hypertext capabilities
\usepackage{multirow}     % in preamble
\usepackage{booktabs}     % optional but improves spacing
\hypersetup{
    colorlinks=true,
    linkcolor=blue,
    filecolor=magenta,
    urlcolor=blue,
    citecolor=blue,
    pdftitle={Quantum Geometric Quadrupole of Cooper Pairs},
    pdfpagemode=FullScreen,
    }
\newcommand{\WQC}[1]{#1}
\newcommand{\JY}[1]{}

\begin{document}
\title{Quantum Geometric Quadrupole of Cooper Pairs}
\author{Wenqin Chen}
\affiliation{Department of Physics, University of Washington, Seattle, Washington 98195, USA}
\author{Kaijie Yang}
\affiliation{Department of Materials Science and Engineering, University of Washington, Seattle, Washington 98195, USA}
\author{Ting Cao}
\affiliation{Department of Materials Science and Engineering, University of Washington, Seattle, Washington 98195, USA}
\author{Shi-Zeng Lin}
\affiliation{Theoretical Division, T-4 and CNLS, Los Alamos National Laboratory, Los Alamos, New Mexico 87545, USA}
\affiliation{Center for Integrated Nanotechnologies (CINT), Los Alamos National Laboratory, Los Alamos, New Mexico 87545, USA}
\author{Jiabin Yu}
\affiliation{Department of Physics and Quantum Theory Project, University of Florida, Gainesville, FL, USA}
\author{Di Xiao}
\affiliation{Department of Materials Science and Engineering, University of Washington, Seattle, Washington 98195, USA}
\affiliation{Department of Physics, University of Washington, Seattle, Washington 98195, USA}
\begin{abstract}
The size of Cooper pairs defines a fundamental length scale of superconductivity, conventionally set by band dispersion and the superconducting gap. This picture breaks down in flat bands, where quenched dispersion makes quantum geometry essential. Here we develop a general framework based on the Cooper pair quadrupole moment, whose trace gives the pair size.
The framework holds for both dispersive and flat-band cases, and provides a unified description of the geometric origin of this length scale.
In particular, when time-reversal symmetry is broken, Berry curvature enters through the phase structure of the pair wavefunction and gives an essential contribution absent from previous quantum-metric theories. Together, Berry curvature and quantum metric impose a geometric lower bound on the pair size. Applying this framework to rhombohedral graphene, we find that the Berry-curvature-induced contribution can dominate and yields pair sizes comparable to experimentally inferred coherence lengths. These results identify Berry curvature as a central geometric ingredient controlling the microscopic length scale of superconductivity.

\end{abstract}
\maketitle

\paragraph*{Introduction.}
%Recent years have seen renewed interest in the Bardeen-Cooper-Schrieffer (BCS) theory~\cite{Bound-Electron-Pairs,Theory-of-Superconductivity} from the perspective of quantum geometry. 
Recent years have seen growing interest in the role of quantum geometry in superconductivity~\cite{torma2022superconductivity,Quantum-metric-and-effective-mass,Revisiting-flat-band-superconductivity,Euler-obstructed-nematic-nodal,Essay-Where-Can-Quantum-Geometry,tian2023evidence,Quantum-geometry-encoded-to-pair,Superconductivity-in-monolayer-FeSe,Quantum-geometric-effect-on-Fulde-Ferrell-Larkin-Ovchinnikov-superconductivity,Quantum-geometry-induced-anapole,Spin-Triplet-Superconductivity-from-Quantum-Geometry-Induced,Enhanced-Kohn-Luttinger,Chiral-Superconductivity-from-Spin-Polarized}.
A notable example is the superfluid weight, which has been shown to receive a significant contribution from quantum geometry, especially when the bands are nearly flat~\cite{peotta2015superfluidity,Geometric-Origin-of-Superfluidity,Band-geometry-Berry-curvature-and-superfluid-weight,Geometric-and-Conventional-Contribution-to-the-Superfluid-Weight,Topology-Bounded-Superfluid-Weight,Superfluid-weight-and-Berezinskii-Kosterlitz-Thouless,Superfluid-Weight-Bounds-from-Symmetry,verma2021optical,Quantum-geometric-superfluid-weight,tanaka2025superfluid}.
This geometric perspective has become especially important in light of recent discoveries of superconductivity in systems with (nearly) flat bands, including twisted graphene moiré systems~\cite{cao2018unconventional,Tuning-superconductivity-in-twisted-bilayer-graphene,lu2019superconductors,codecido2019correlated,saito2020independent,stepanov2020untying,oh2021evidence,park2021tunable}, rhombohedral graphene~\cite{chen2019signatures,zhou2021superconductivity,han2025signatures,patterson2025superconductivity,yang2025impact,choi2025superconductivity,kumar2025superconductivity}, and twisted transition metal dichalcogenides~\cite{jia2025anomalous,xia2025superconductivity,guo2025superconductivity}.

Beyond the superfluid weight, another fundamental characteristic of a superconductor is the size of its Cooper pairs.
In conventional BCS theory, the Cooper pair size is set by band dispersion and the superconducting gap~\cite{Bound-Electron-Pairs,Theory-of-Superconductivity}. 
This description breaks down in (nearly) flat bands, where the dispersion is quenched and quantum geometry becomes essential.
Recent works have shown that the Cooper pair size also receives a contribution from the quantum metric, the symmetric part of the quantum geometric tensor~\cite{Extracting-quantum-geometric-effects,Ginzburg-Landau-Theory-of-Flat-Band,Coherence-length-and-quantum-geometry,Pair-size-and-quantum-geometry,Correlation-functions-and-characteristic-lengthscales,hu2025anomalous,li2025vortexstatescoherencelengths,elden2026correlationlengthsflatbandsuperconductivity}.  However, these studies have largely focused on time-reversal-symmetric settings, leaving open the role of the Berry curvature, the antisymmetric part of the quantum geometric tensor~\cite{Berry-phase-effects-on-electronic-properties}.
This omission is notable because Berry curvature can strongly affect two-body bound states in Bloch bands.
In excitons, for example, Berry curvature modifies the relative motion of electron-hole pairs, especially in higher-angular-momentum channels~\cite{Berry-Phase-Modification-to-the-Energy,Signatures-of-Bloch-Band-Geometry}.
Motivated by recent reports of higher-angular-momentum superconductivity in rhombohedral graphene, where Berry curvature is large near the Fermi surface~\cite{han2025signatures}, we ask whether Berry curvature can also contribute to the Cooper pair size.

\begin{figure}[b]
    \includegraphics[width=0.85\linewidth]{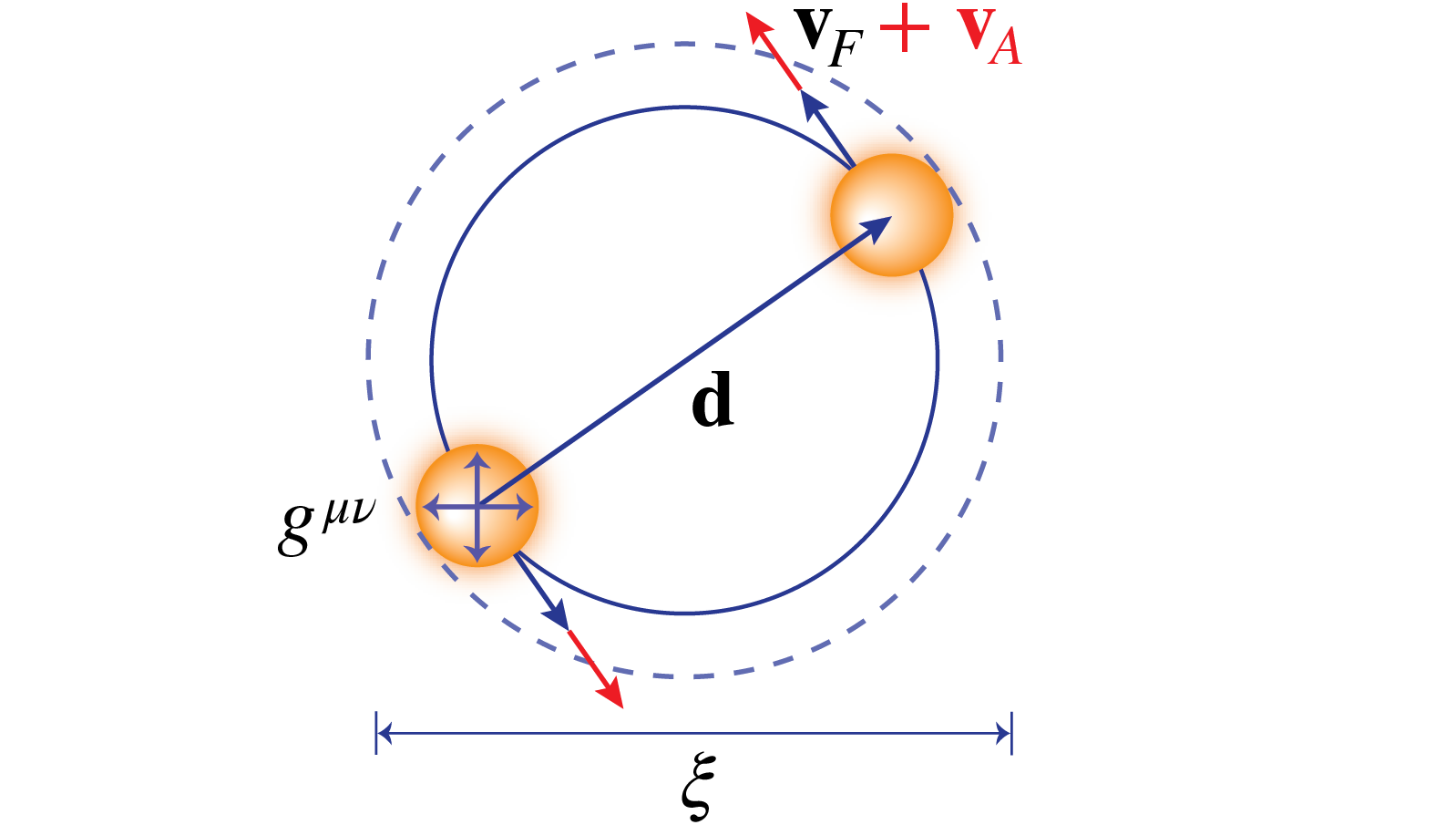}
    \caption{Semiclassical picture of a two-body bound state in Bloch bands. Under an attractive interaction $V$, two electrons orbit each other with velocity $\mathbf v_F+\mathbf v_A$, where the anomalous velocity $\mathbf v_A=\nabla V\times\boldsymbol\Omega$ arises from Berry curvature $\boldsymbol\Omega$. The separation $\mathbf d$ between two electron centers sets the size of the circular trajectory, while each electron’s intrinsic spread is characterized by the quantum metric $g^{\mu\nu}$. Together, these two contributions determine the Cooper pair size. }
    \label{fig: two-body}
\end{figure}

In this work, we resolve this question by developing a general theory of the Cooper pair quadrupole moment, whose trace defines the pair size. For intraband pairing, we show that once time-reversal symmetry (TRS) is broken, the Cooper pair size acquires an essential geometric contribution from Berry curvature---absent in previous theories---which enters through the phase structure of the pair wavefunction.  Together with the quantum metric, this contribution sets a geometric lower bound on the Cooper pair size. 
Applying our theory to rhombohedral graphene with spin-polarized intravalley pairing, we find a substantial Berry-curvature-induced phase contribution, ranging from about 50\% to nearly 100\% of the total pair size.
The resulting Cooper pair size is comparable in magnitude to the experimentally observed coherence length, suggesting that quantum geometry sets the characteristic microscopic length scale of superconductivity in rhombohedral graphene. Our theory thus identifies a concrete and previously missing consequence of Berry curvature in superconductors, and establishes quantum geometry as a central ingredient controlling the microscopic structure of Cooper pairs.

\paragraph*{Quadrupole moment of Cooper pairs.}
To motivate the general formalism, we begin with a semiclassical picture of the Cooper pair quadrupole moment, as illustrated in Fig.~\ref{fig: two-body}. When two Bloch electrons form a bound state via an attractive interaction $V$, their charge centers orbit around each other with a transverse velocity, which consists of a Fermi velocity contribution $\vb v_F$ from the band dispersion and an anomalous velocity $\vb v_A = \nabla V \times \bm \Omega$ from the Berry curvature \cite{Berry-phase-effects-on-electronic-properties}.  Importantly, the anomalous velocity can drive relative motion even in a flat band, where the dispersion is quenched~\cite{Berry-Phase-Modification-to-the-Energy,Signatures-of-Bloch-Band-Geometry}. This relative motion separates the two charge centers by a vector $\vb d$, and the tensor $\vb d\otimes\vb d$ characterizes the spatial extent of the circular motion. In addition, each Bloch electron wavepacket carries an intrinsic gauge-invariant Wannier spread around its center, quantified by the quantum metric~\cite{Maximally-localized-generalized-Wannier}. 
The Cooper pair quadrupole moment (whose trace defines the pair size) is therefore controlled by both the relative motion of the pair and the intrinsic spread of its constituents, set respectively by the Berry curvature, band dispersion, and quantum metric.

Having established the physical picture, we now formulate the problem within a band-projected BCS framework.  For simplicity, we consider pairing in one band without spin degeneracy.
The Cooper pair state is 
\begin{equation}
\label{eq: Cooper pair state}
\ket{\Psi}=\sum_{\vb k}\psi_{\vb k}\ket{\vb k_P}, \quad
\ket{\vb k_P}=\hat a^\dagger_{\vb k}\hat a^\dagger_{-\vb k}\ket{\text{GS}},
\end{equation}
where \(\hat a^\dagger_{\vb k}\) creates an electron in the Bloch state with periodic part $\ket{u_{\boldsymbol{k}}}$, $\ket{\vb k_P}$ is the pair basis state, and $\vb k$ is the relative momentum. The envelope function \(\psi_{\vb k}\) is determined by solving the two-body Schrödinger equation for \(\hat H = \hat h + \hat V\), where \(\hat h\) is the single-particle Hamiltonian and $\hat V$ is the pairing interaction projected onto the active band.

\begin{table}
    \caption{\label{table: geometric quantities} Two-body quantum geometry. Here, $\phi_{\vb k} = \arg \psi_{\vb k}$ is the phase of the envelope function, and $A^{\mu}_{\vb k}$, $\Omega_{\vb k}$, $g^{\mu\nu}$ denote the single-particle Berry connection, Berry curvature, and quantum metric, respectively. $A^\mu_P$ is gauge dependent, ${D}_{k_\mu}$ is gauge covariant, and $\Omega_P$, $g^{\mu\nu}_P$, $d^\mu_G$, $\mathcal Q^{\mu\nu}_G$ are gauge-invariant quantities.}
    \renewcommand{\arraystretch}{1.3}
\begin{ruledtabular}
\begin{tabular}{l l }
Pair geometric quantities & Expressions\\[1.2pt]
\hline
   Berry connection & $A_P^\mu(\vb k) =  A^\mu_{\vb{k}} - {A}^\mu_{-\vb{k}}$  \\
   Berry curvature  & $\Omega_P(\vb k) =  \Omega_{\vb k}+  \Omega_{-\vb k}$   \\
   Quantum metric  & $g^{\mu\nu}_P(\vb k) = g^{\mu\nu}_{\vb k} + g^{\mu\nu}_{-\vb k}$   \\
   Covariant derivative & ${D}_{k_\mu} = i\partial_{k_\mu} + A^\mu_P(\vb k)$\\
   Geometric dipole  & $d^\mu_G(\vb k) = \partial_{k_\mu}\phi_{\vb k} - {A}^\mu_P(\vb k)$  \\
   Geometric quadrupole & $\mathcal{Q}^{\mu\nu}_G(\vb k) = d^\mu_G(\vb k) d^\nu_G(\vb k)  + g^{\mu\nu}_P(\vb k)$
   \end{tabular}
\end{ruledtabular}
\end{table}

%Fourier transforming the envelope function gives the real-space pair wavefunction, $\ket{\psi(\vb r)}=\sum_{\vb k}\psi_{\vb k}\,e^{i\vb k\cdot\vb r}\ket{u_{\vb k}}\otimes\ket{u_{-\vb k}}$

Following the primitive quadrupole of Ref.~\cite{Haldane_2023_Quadrupole}, we define the Cooper pair quadrupole moment as~\cite{SM}, $
Q^{\mu\nu}=\bra{\Psi}\hat r^\mu\hat r^\nu\ket{\Psi}
$, where $\vb r$ is the \emph{relative} coordinate.
Converting back to momentum space, the $\vb r$ operator becomes a derivative with respect to $\vb k$ \cite{SM}, yielding
\begin{equation}
\label{eq: raw quadrupole}
\begin{aligned}
Q^{\mu\nu} =-\lim_{\vb q \to 0}\sum_{\vb k}
\partial_{q_\mu}\partial_{q_\nu}
\qty(
\psi_{\vb k+\tfrac{\vb q}{2}}
\psi^*_{\vb k-\tfrac{\vb q}{2}}
\Lambda^P_{\vb k-\frac{\vb q}{2},\,\vb k+\frac{\vb q}{2}}).
\end{aligned}
\end{equation}
Here,
\(\Lambda^{P}_{\vb k,\vb k'}=
\langle u_{\vb k}|u_{\vb k'}\rangle
\langle u_{-\vb k}|u_{-\vb k'}\rangle\)
is the pair form factor built from single-particle Bloch overlaps. Under the single-particle gauge transformation
\(\ket{u_{\vb k}}\to e^{i\alpha_{\vb k}}\ket{u_{\vb k}}\), $\Lambda^{P}_{\vb k,\vb k'}$ acquires a phase
\(e^{i(\alpha_{\vb k'}-\alpha_{\vb k}+\alpha_{-\vb k'}-\alpha_{-\vb k})} \).
Meanwhile, the pair basis transforms as
\(\ket{\vb k_P}\to e^{i(\alpha_{\vb k}+\alpha_{-\vb k})}\ket{\vb k_P}\), so gauge invariance of the physical pair state requires
\(\psi_{\vb k}\to e^{-i(\alpha_{\vb k}+\alpha_{-\vb k})}\psi_{\vb k}\).
These phases cancel so \(Q^{\mu\nu}\) is gauge invariant.

The overlaps $\langle u_{\vb k}|u_{\vb k'}\rangle$ encode the underlying quantum geometry, defined by the quantum geometric tensor \cite{Provost1980,The-insulating-state-of-matter}, \(\mathcal{G}^{\mu\nu}_{\vb k} = \bra{\partial_{k_\mu} u_{\vb k}}P_{\vb k} \ket{\partial_{k_\nu} u_{\vb k}}\), where $P_{\vb k} = 1 - \ket{u_{\vb k}}\bra{u_{\vb k}}$. Its symmetric and antisymmetric parts give, respectively, the quantum metric and Berry curvature via $\mathcal{G}^{\mu\nu}_{\vb k} = g^{\mu\nu}_{\vb k} + \tfrac{i}{2}\epsilon^{\mu\nu}\Omega_{\vb k}$ \cite{yu2025quantum}.
To second order in $\vb q$, the Bloch overlap takes the compact exponential form \(\exp(-iA^\mu_{\vb k} q_\mu\!-\!\tfrac12 g^{\mu\nu}_{\vb k}q_\mu q_\nu) \),
where \(A^\mu_{\vb k}=i\langle u_{\vb k}|\partial_{k_\mu}u_{\vb k}\rangle\) is the Berry connection.
This form reproduces the exact derivatives up to second order as required in Eq.\,(\ref{eq: raw quadrupole}), while encapsulating both geometric ingredients—Berry phase (through $A^\mu_{\vb k}$) and quantum metric $g^{\mu\nu}_{\vb k}$—in a single exponential kernel.

The pair form factor then follows simply as the product of two such single-particle overlaps,
\begin{equation}
\label{eq: form factor}
    \begin{aligned}
        \Lambda^{P}_{\vb{k}-\frac{\vb q}{2},\,\vb{k}+\frac{\vb q}{2}}
=
&\exp\left[-iA_{P}^\mu(\vb k) q_\mu-\frac12 g_{P}^{\mu\nu}(\vb k)q_\mu q_\nu\right]\\
&+\mathcal O(q^3),
    \end{aligned}
\end{equation}
where the pair Berry connection \(A_P^\mu(\vb k)=A^\mu_{\vb k}-A^\mu_{-\vb k}\) and the pair quantum metric
\(g_P^{\mu\nu}(\vb k)=g^{\mu\nu}_{\vb k}+g^{\mu\nu}_{-\vb k}\)
are defined in Table~\ref{table: geometric quantities}.
The resulting pair Berry curvature
\(\Omega_P(\vb k)=\nabla_{\vb k}\times\vb A_P(\vb k)\)
is the sum of the two single-particle curvatures \cite{Berry-Phase-in-Magnetic-Superconductors, Topological-Nodal-Cooper}. Clearly, pair Berry curvature effects can arise only when the normal-state TRS is broken, since otherwise $\Omega_{\vb k}$ and $\Omega_{-\vb k}$ cancel completely.

It now becomes straightforward to evaluate the pair quadrupole moment, since Eq.~(\ref{eq: form factor}) allows the derivatives in Eq.~(\ref{eq: raw quadrupole}) to be carried out exactly (only $\partial^2_q$ at $\vb q=0$ contributes). 
The detailed derivation is provided in the Supplemental Material (SM)~\cite{SM}.
The resulting expression can be decomposed as
\begin{equation}
\label{eq: pair quadrupole moment}
    Q^{\mu\nu} = Q^{\mu\nu}_G + Q_A^{\mu\nu} \;.
\end{equation}
The first term is the geometric contribution, given by
\begin{equation}
\begin{aligned}
&Q_G^{\mu\nu} = \sum_{\vb k} |\psi_{\vb k}|^2 \mathcal{Q}_G^{\mu\nu}(\vb k),\\
&\mathcal{Q}^{\mu\nu}_{\mathrm{G}}(\vb k)
= [\partial_{k_\mu}\phi_{\vb k}-A^{\mu}_{P}(\vb k)]
   [\partial_{k_\nu}\phi_{\vb k}-A^{\nu}_{P}(\vb k)]
   + g^{\mu\nu}_{P}(\vb k).
\end{aligned}
\label{eq: geometric quadrupole}
\end{equation}
The $\vb k$-resolved quantity $\mathcal{Q}^{\mu\nu}_{\mathrm G}(\vb k)$ defines the \textit{pair geometric quadrupole}. 
Here $d^{\mu}_{G}(\vb k) = \partial_{{k}_{\mu}}\phi_{\vb k} - A^{\mu}_{P}(\vb k)$, which we refer to as pair geometric dipole, is the gauge-invariant combination identified in Ref.~\cite{Topological-Nodal-Cooper}, and $\phi_{\vb k} = \arg \psi_{\vb k}$ is the phase of the envelope function.
The rank-2 tensor $\mathcal{Q}^{\mu\nu}_{\mathrm G}(\vb k)$ has the structure dipole $\otimes$ dipole $+$ metric, and can be understood intuitively from Fig.~\ref{fig: two-body}: the first term arises from the dipole separation between the charge centers, while the second term accounts for the intrinsic spread of each constituent electron.
The second contribution arises from the amplitude variation of the envelope function,
\begin{equation}
    Q^{\mu\nu}_A = \sum_{\vb k}\frac{1}{2}\qty(
\partial_{k_\mu}|\psi_{\vb k}|\,\partial_{k_\nu}|\psi_{\vb k}|
-|\psi_{\vb k}|\,\partial_{k_\mu}\partial_{k_\nu}|\psi_{\vb k}|
).
\label{eq: amplitude contribution}
\end{equation}

The square root of the trace of $Q^{\mu\nu}$ defines the Cooper pair size,
\begin{equation}
\label{eq: pair size}
    \xi_{\rm Pair} = \sqrt{\Tr\,Q^{\mu\nu}}= \sqrt{\xi^2_{\rm Amp} + \xi^2_{\rm Metric} + \xi^2_{\rm Phase}}\,,
\end{equation}
where the three contributions arise respectively from the amplitude variation, quantum metric, and phase structure:
\begin{equation}
    \begin{aligned}
        &\xi^2_{\rm Amp} = \sum_{\vb k}\frac{1}{2}\left[\qty(\nabla_{\vb k}|\psi_{\vb k}|)^2 - |\psi_{\vb k}|\nabla^2_{\vb k}|\psi_{\vb k}|\right],\\
        &\xi^2_{\rm Metric} = \sum_{\vb k}|\psi_{\vb k}|^2 \Tr [g_P(\vb k)],\\
        &\xi^2_{\rm Phase} = \sum_{\vb k}|\psi_{\vb k}|^2\left[\nabla_{\vb k}\phi_{\vb k}-\vb A_{P}(\vb k)\right]^2.
    \end{aligned}
    \label{eq: pair size contributions}
\end{equation}
Here \(\phi_{\vb k}=\arg\psi_{\vb k}\), and \(\mathrm{Tr}[g_P(\vb k)]=\delta_{\mu\nu}g_P^{\mu\nu}(\vb k)\) is the trace of the pair quantum metric (more mathematically precise, the Dirichlet functional/energy~\cite{Yu_2025_WL_ideal}).

Having established the decomposition of $\xi_{\rm Pair}$, we now examine each contribution in detail.  The amplitude term $\xi_{\rm Amp}$ dominates in conventional superconductors, where quantum geometric effects are negligible compared with the dispersion and TRS is preserved. In this limit, $\psi_{\vb k}$ can be chosen real and is determined by the dispersion and pairing strength. Accordingly, $\xi_{\rm Amp}$ reduces to the BCS relation $\xi \sim \hbar v_F/E_B$, with $v_F$ the Fermi velocity and $E_B$ the binding energy~\cite{deGennes1999}.

The metric term $\xi_{\rm Metric}$ corresponds to the quantum metric contribution discussed in Refs.~\cite{Extracting-quantum-geometric-effects,Ginzburg-Landau-Theory-of-Flat-Band,Coherence-length-and-quantum-geometry,Pair-size-and-quantum-geometry,Correlation-functions-and-characteristic-lengthscales,hu2025anomalous,li2025vortexstatescoherencelengths,elden2026correlationlengthsflatbandsuperconductivity}. When $\Tr[g_P(\vb k)]$ is approximately constant over the support of the normalized envelope function, the metric contribution reduces to this constant, $\xi^2_{\rm Metric}\simeq \Tr[g_P]$, since $\sum_{\vb k}|\psi_{\vb k}|^2=1$. More generally, $\xi^2_{\rm Metric}$ is the envelope-weighted average of the pair quantum metric trace. For a generic Cooper pair, the total pair size contains all three contributions in Eq.~(\ref{eq: pair size contributions}).

The phase term $\xi_{\rm Phase}$ is also gauge invariant. Using the gauge transformation defined after Eq.~(\ref{eq: raw quadrupole}), the envelope phase and pair Berry connection transform as
\(\phi_{\vb k} \to \phi_{\vb k} - \alpha_{\vb k} - \alpha_{-\vb k}\) and
\(\vb A_P(\vb k) \to \vb A_P(\vb k) - \nabla_{\vb{k}}(\alpha_{\vb k} + \alpha_{-\vb k})\),
so that 
\(\vb d_G(\vb k) = \nabla_{\vb k} \phi_{\vb k} - \vb A_P(\vb k)\)
is invariant~\cite{Topological-Nodal-Cooper,Berry-Phase-in-Magnetic-Superconductors}. 
Since $\xi_{\rm Amp}$ and $\xi_{\rm Metric}$ are also gauge invariant, the total pair size $\xi_{\rm Pair}$ in Eq.~(\ref{eq: pair size}) is a gauge-invariant observable.

\paragraph*{Geometric bound on Cooper pair size.}
We now show that the Cooper pair size satisfies a lower bound set by two-body quantum geometry.
Starting from the amplitude contribution in Eq.~(\ref{eq: amplitude contribution}), we integrate by parts over the Brillouin zone, so that the boundary term vanishes. Combining it with the geometric contribution in Eq.~(\ref{eq: geometric quadrupole}), we obtain the pair quadrupole operator
\begin{equation}
    \hat r^\mu\hat r^\nu
=\sum_{\vb k}\ket{\vb k_P}
\left[
 D^\dagger_{k_\mu} D_{k_\nu}
+g^{\mu\nu}_P(\vb k)
\right]
\bra{\vb k_P},
\end{equation}
where ${D}_{k_\mu} = i\partial_{k_\mu} + A_P^\mu(\vb k)$ is the covariant derivative for the relative motion. The trace gives the $\hat r^2$ operator projected onto the pair basis. The covariant-derivative term contains the amplitude and phase-gradient contributions, $\xi^2_{\rm Amp}$ and $\xi^2_{\rm Phase}$ in Eq.~(\ref{eq: pair size contributions}).

We next focus on the case where the envelope function is supported, or strongly concentrated, in a momentum-space region $\mathcal R$ over which the pair Berry curvature does not change sign. 
For definiteness, we take $\Omega_P(\vb k)\ge0$ in $\mathcal R$. 
Introducing the holomorphic operators $D=(D_{k_x}+iD_{k_y})/2$ and $\Bar D=(D_{k_x}-iD_{k_y})/2$ and completing the square, we obtain \cite{SM}
\begin{equation}
    \hat r^2
=\sum_{\vb k}\ket{\vb k_P} \,
    \Bigl(\,4\Bar DD
+\Omega_P(\vb k)
+\mathrm{Tr}[g_P(\vb k)]\,\Bigr)\bra{\vb k_P},
\end{equation}
where we have used $\left[D_{k_x},D_{k_y}\right]=i\Omega_P(\vb k)$.
If the curvature has the opposite sign, the same argument follows by interchanging $D$ and $\Bar D$. 
The single-particle trace inequality \cite{roy2014band} immediately generalizes to the pair quantities, $\mathrm{Tr}[g_P(\vb k)]\ge|\Omega_P(\vb k)|$.

\begin{figure*}
\includegraphics[width=0.9\linewidth]{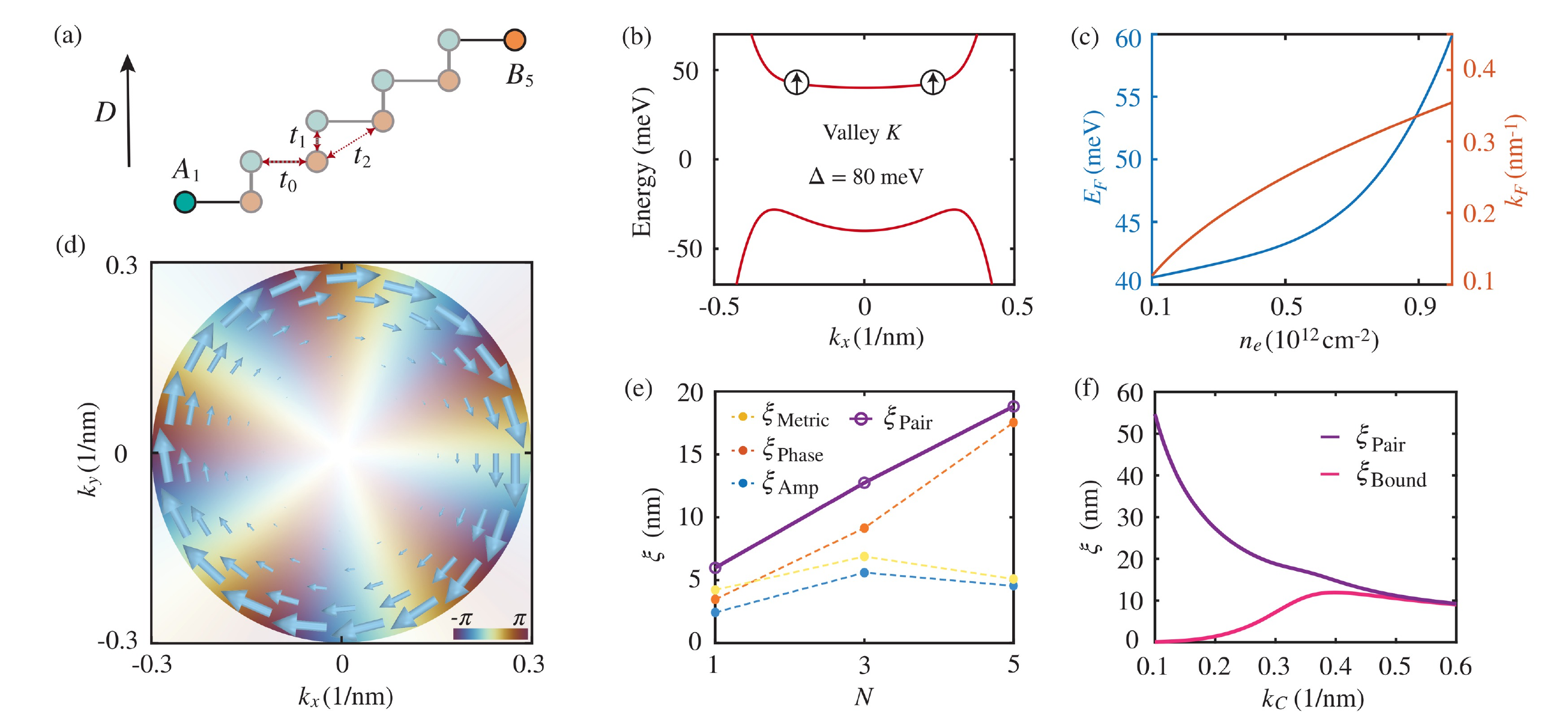}
    \caption{Cooper pair size in the two-band model of rhombohedral graphene. (a) Schematic of the two-band model formed by the two surface orbitals under a displacement field. (b) Low-energy band dispersion for the potential difference $\Delta = 80$ meV, which is tunable by the displacement field. (c) Fermi level and Fermi momentum as functions of the electron density. (d) Phase vortex structure (color) of the envelope function and the pair geometric dipole $\vb d_G$ (arrows). (e) Cooper pair size and its contributions as functions of the winding number $N$, which is varied as a model parameter. Here $k_C = 0.3\,\rm nm^{-1}$. (f) Cooper pair size and its geometric lower bound as functions of the momentum cutoff $k_C$. Here $N = 5$. }\label{fig:rhombohedral graphene}
\end{figure*}

Since $\Bar DD$ is positive semidefinite, the expectation value of $\hat r^2$ obeys
\begin{equation}
\label{eq: bound}
    \langle \hat r^2\rangle \ge 
    \sum_{\vb k\in\mathcal R} |\psi_{\vb k}|^2
    \Bigl(
    \Omega_P(\vb k) + \Tr[g_P(\vb k)]
    \Bigr)
    \equiv \xi^2_{\rm Bound}.
\end{equation}
Applying the pair trace inequality $\mathrm{Tr}[g_P(\vb k)]\ge\Omega_P(\vb k)$, we further obtain $\langle \hat r^2\rangle
    \ge
    2\sum_{\vb k\in\mathcal R}
    |\psi_{\vb k}|^2\Omega_P(\vb k)$, a lower bound determined solely by the envelope-weighted average of pair Berry curvature. 
Equation~(\ref{eq: bound}) is therefore a geometric lower bound: even when band dispersion is quenched, the Cooper pair size cannot be made arbitrarily small, because the pair Berry curvature and pair quantum metric set an intrinsic length scale.

\paragraph*{Rhombohedral graphene.}
As an illustrative example, we now apply our theory to the two-band model that captures the low-energy physics of rhombohedral graphene~\cite{koshino2009trigonal,geier2025chiral,Family-of-multilayer}. The low-energy effective Hamiltonian for rhombohedral pentalayer graphene is adapted from Refs.~\cite{koshino2009trigonal,Berry-curvature-of-low-energy-excitons}
\begin{equation}
    \begin{aligned}
\label{eq:R5G hamiltonian}
\hat h
=\begin{pmatrix}
\frac{\Delta}{2}+\gamma_{-}(k_x^2+k_y^2)
&
\dfrac{v_F^{5}}{t_1^{4}}(k_x-i k_y)^{5}
\\[6pt]
\dfrac{v_F^{5}}{t_1^{4}}(k_x+i k_y)^{5}
&
-\frac{\Delta}{2}+\gamma_{+}(k_x^2+k_y^2)
\end{pmatrix},
    \end{aligned}
\end{equation}
written in the basis of the two surface orbitals $(A_1,B_5)$ shown in Fig.~\ref{fig:rhombohedral graphene}(a). For the numerical results we use $a=0.246\,\rm nm$, $t_0=3.1\,\rm eV$, $t_1=380\,\rm meV$, and $t_2=141\,\rm meV$, with $v_F=(\sqrt{3}/2)t_0a$, $v_2=(\sqrt{3}/2)t_2a$, and $\gamma_{\pm}\simeq 2v_Fv_2/t_1\pm \Delta v_F^2/(4t_1^2)$. The displacement-field-induced gap $\Delta$ is set to $\Delta=80\,\rm meV$. The band dispersion and Fermi level are shown in Fig.~\ref{fig:rhombohedral graphene}(b) and (c).

Motivated by recent reports of chiral superconductivity in rhombohedral graphene, we consider spin-polarized intravalley pairing in the conduction band. For an analytic treatment, we assume a local attractive interaction and restrict pairing to $|\vb k|<k_C$, with $k_C$ of order $k_F$ \cite{chou2025intravalley}, so that the dispersion can be neglected. The two-body Schr\"odinger equation then becomes
\begin{equation}
    E_B\,\psi(\vb{k}) = \frac{1}{2S}\sum_{|\vb{k}'|<k_C} V_{\vb k,\vb k'}\,\psi(\vb{k}'),
\end{equation}
where $V_{\vb k,\vb k'} = -U(\Lambda^P_{\vb k,\vb k'} - \Lambda^P_{-\vb k,\vb k'})/2$ is the antisymmetrized band-projected interaction \cite{Family-of-multilayer,From-Fractionalization-to-Chiral}; here $\Lambda^P_{\vb k,\vb k'}$ is the pair form factor introduced after Eq.~(\ref{eq: raw quadrupole}) and $S$ is the system area. In this limit $V_{\vb k,\vb k'}$ is separable, and the envelope function can be obtained analytically as
\begin{equation}
\label{eq: envelope function}
    \psi_{\vb{k}} = \frac{1}{\sqrt{\mathcal{N}}}\frac{(k_x - ik_y)^5}{f(\vb{k})},
\end{equation}
where $\mathcal{N}$ is a normalization factor and $f(\vb k)$ is a real-valued function given in the SM \cite{SM}. Figure~\ref{fig:rhombohedral graphene}(d) shows the phase texture of $\psi_{\vb k}$, whose phase winds by five around the valley center. The corresponding pair geometric dipole $\vb d_G(\vb k)$ [Table.~\ref{table: geometric quantities}] exhibits a vortex characterized by the circulation $\oint d\vb k\!\cdot\!\vb d_G(\vb k)$.

We now evaluate the three contributions to the Cooper pair size in Eq.~(\ref{eq: pair size contributions}); the results are shown in Fig.~\ref{fig:rhombohedral graphene}(e). At winding $N=5$, the phase contribution $\xi_{\rm Phase}$ dominates. This is transparent from Eq.~(\ref{eq: envelope function}): the factor $(k_x-ik_y)^5$ gives $\phi_{\vb k}=-5\theta_{\vb k}$, producing the fivefold vortex shown in Fig.~\ref{fig:rhombohedral graphene}(d). The phase winding and nodal structure of this momentum-space vortex are analogous to those of an Abrikosov vortex in real space under a magnetic field. Although $\nabla_{\vb k}\phi_{\vb k}$ is singular at the vortex core, $|\psi_{\vb k}|$ vanishes as $k^5$, so the weighted integrand $|\psi_{\vb k}|^2 d_G^2(\vb k)$ remains finite. Such a vortex can make a significant contribution to $\xi_{\rm Phase}$. Indeed, treating $N$ as a model parameter (physically $N=5$) in Fig.~\ref{fig:rhombohedral graphene}(e) shows that $\xi_{\rm Phase}$ grows with $N$ and becomes the leading length scale for $N\geq 5$. Figure~\ref{fig:rhombohedral graphene}(f) further shows $\xi_{\rm Pair}$ and $\xi_{\rm Bound}$ as functions of $k_C$. The bound provides an accurate estimate at larger $k_C$, but becomes less accurate at smaller $k_C$.

The decomposition also allows us to isolate the quantum metric contribution emphasized in Refs.~\cite{Extracting-quantum-geometric-effects,Ginzburg-Landau-Theory-of-Flat-Band,Coherence-length-and-quantum-geometry,Pair-size-and-quantum-geometry,Correlation-functions-and-characteristic-lengthscales,hu2025anomalous,li2025vortexstatescoherencelengths,elden2026correlationlengthsflatbandsuperconductivity}. Retaining only the quantum metric term gives a length scale of a few nanometers for the parameters of Fig.~\ref{fig:rhombohedral graphene}, well below the experimental coherence length. In this TRS-broken setting, the Berry-curvature-induced phase contribution is therefore essential and must be included in a complete theory.

Using parameters relevant to the experiment in rhombohedral graphene \cite{han2025signatures}, the resulting pair size is of the same order as the coherence length inferred experimentally from the upper critical field, which is on the order of tens of nanometers.
Although our theory computes a microscopic pair size whereas the experiment probes a phenomenological Ginzburg-Landau coherence length, the order-of-magnitude agreement is nevertheless encouraging, especially because the observed superconducting states lie on the BCS side of the BCS-BEC crossover~\cite{The-BCS-BEC-crossover}. 

We have also applied our theory to the Berry Trashcan model \cite{bernevig2025berrytrashcanmodelinteracting,li2025berrytrashcanshortrange} and found similar orders of magnitude and trends, as summarized in the SM \cite{SM}.

\paragraph*{Conclusion and Discussion.}
In summary, we establish the geometric quadrupole moment as a fundamental quantity linking quantum geometry to the microscopic structure of Cooper pairs. 
The pair quadrupole moment directly enters the London equation through a nonlocal Pippard/BCS kernel that accounts for the finite spread of the pair \cite{Theory-of-Superconductivity,Pippard1953_FieldCurrent}, making it accessible via nonlocal electromagnetic response measurements \cite{Direct-Observation-of-Nonlocal,Observation-of-nonexponential,Nonlocal-effect-and-dimensions}. 
Similar probes in flat-band superconductors could thus provide a direct window into the geometric structure of Cooper pairs predicted here. 
Our framework applies to generic two-body bound states in Bloch bands and can be straightforwardly extended to systems such as excitons.

\paragraph*{Acknowledgment.}
We thank Xiaodong Hu, Ammar Jahin, Yafei Ren, Enrico Rossi, Mark Rudner, and Jihang Zhu for helpful discussion.
This work is supported by the University of Washington Molecular Engineering Materials Center (MEM-C, NSF grant DMR-2308979).
J. Y.'s work is supported by startup funds at University of Florida.
SZL is partially supported by the U.S. Department of Energy (DOE) National Nuclear Security Administration (NNSA) under Contract No. 89233218CNA000001 through the Laboratory Directed Research and Development (LDRD) Program and was performed, in part, at the Center for Integrated Nanotechnologies, an Office of Science User Facility operated for the DOE Office of Science, under user Proposals No. 2018BU0010 and No. 2018BU0083.
This research used resources of the National Energy Research Scientific Computing Center, a DOE Office of Science User Facility supported by the Office of Science of the U.S. Department of Energy under Contract No. DE-AC02-05CH11231 using NERSC award BES-ERCAP0037097 and BES-ERCAP0037104.
This work was also facilitated through the use of advanced computational, storage, and networking infrastructure provided by the Hyak supercomputer system and funded by the University of Washington Molecular Engineering Materials Center at the University of Washington (DMR-2308979)

\onecolumngrid
\clearpage

\setcounter{equation}{0}
\setcounter{figure}{0}
\setcounter{section}{0}
\renewcommand{\theequation}{S\arabic{equation}}
\renewcommand{\thefigure}{S\arabic{figure}}
\renewcommand{\thesection}{S\arabic{section}}
\renewcommand{\theHequation}{S\arabic{equation}}
\renewcommand{\theHfigure}{S\arabic{figure}}
\renewcommand{\theHsection}{S\arabic{section}}

\begin{center}
\textbf{\large Supplemental Materials for ``Quantum Geometric Quadrupole of Cooper Pairs''}
\end{center}
\vspace{0.5em}

\tableofcontents

\section{Derivations of Cooper pair quadrupole moment}
To compute the Cooper pair quadrupole, we first transform the Cooper pair creation operator $\mathcal{P}^\dagger = \sum_{\vb{k}} \psi(\vb{k})a^\dagger_{\vb{k}\uparrow}a^\dagger_{-\vb{k}\downarrow}$ to real space, by transforming the field operators as,
\begin{equation}
\begin{aligned}
    a_{\vb{k}\uparrow} = \int d^2x \sum_{\alpha} e^{-i\vb{k}\cdot \vb{x}} u^*_{\alpha\vb{x}}(\vb{k}) c_{\vb{x}\alpha\uparrow}.
\end{aligned}
\end{equation}
The Cooper pair creation operator in real space then becomes
\begin{equation}
\begin{aligned}
    \mathcal{P}^\dagger = &\int d^2x_1 d^2x_2 \sum_{\alpha\beta} \sum_{\vb{k}} \psi(\vb{k}) e^{i\vb{k}\cdot \vb{r}} u_{\alpha\vb{x}_1}(\vb{k}) u_{\beta\vb{x}_2}(-\vb{k}) c^\dagger_{\vb{x}_1\alpha\uparrow} c^\dagger_{\vb{x}_2\beta\downarrow}\\
    = &\int d^2x_1 d^2x_2 \sum_{\alpha\beta} \psi_{\alpha\beta}(\vb{x}_1,\vb{x}_2) c^\dagger_{\vb{x}_1\alpha\uparrow} c^\dagger_{\vb{x}_2\beta\downarrow},
\end{aligned}
\end{equation}
where $\vb{r} = \vb{x}_1 - \vb{x}_2$ is the relative coordinate. \WQC{We assume that the pair envelope function has no dependence on the center-of-mass momentum.} The real-space Cooper pair wavefunction is given by
\begin{equation}
\begin{aligned}
    \psi_{\alpha\beta}(\vb{x}_1,\vb{x}_2) = \sum_{\vb{k}} \psi(\vb{k}) e^{i\vb{k}\cdot \vb{r}} u_{\alpha\vb{x}_1}(\vb{k}) u_{\beta\vb{x}_2}(-\vb{k}).
\end{aligned}
\end{equation}
where $\psi(\vb{k})$ is the Cooper pair envelope function in momentum space. The pair state is defined as
\begin{equation}
\begin{aligned}
    \ket{\Psi} = \mathcal{P}^\dagger \ket{GS}
\end{aligned}
\end{equation}
and the normalization is 
\begin{equation}
\begin{aligned}
    \mathcal{N} = \bra{GS}\mathcal{P} \mathcal{P}^\dagger \ket{GS} = \int d^2x_1 d^2x_2 \sum_{\alpha\beta} |\psi_{\alpha\beta}(\vb{x}_1,\vb{x}_2)|^2.
\end{aligned}
\label{normalization}
\end{equation}
where $\ket{GS}$ is the ground state.

%\JY{Let's put a hat on the operator as the non-operator quantities are used before in the FT. }
\WQC{We use the definition of the primitive electric quadrupole tensor defined in Ref.~\cite{Haldane_2023_Quadrupole},
\begin{equation}
    \hat q^{\mu\nu} = \frac{1}{2} \sum_i q_i (\hat x^{\mu}_i - \hat{\bar x}^{\mu}) (\hat x^{\nu}_i - \bar x^{\nu}),
\end{equation}
where $\Bar{\vb x}$ is the charge center given by
\begin{equation}
    \sum_i q_i\, \hat{\vb x}_i = \hat{\Bar{\vb x}}\sum_i q_i.
\end{equation}
Now we write this definition in two-body way as $\ket{\Psi}$ is a two-body state, where $i = 1,2$ labels two electrons and $q_i =q$ denotes equal charges.
The charge center becomes
\begin{equation}
    \hat{\Bar{\vb x}} = \frac{q(\hat{\vb x}_1 + \hat{\vb x}_2)}{2q} = \frac{1}{2}(\hat{\vb x}_1 + \hat{\vb x}_2)\equiv \hat{\vb R}
\end{equation}
which is exactly the center-of-mass coordinate $\hat{\vb R}$. Plugging $\hat{\bar{\vb x}}$ into $\hat q^{\mu\nu}$, we have
\begin{equation}
    \hat{\vb x}_1 - \hat{\bar{\vb x}} = \frac{\hat{\vb r}}{2},\qquad \hat{\vb x}_2 - \hat{\bar{\vb x}} = -\frac{\hat{\vb r}}{2}
\end{equation}
where we have defined the relative coordinate $\hat{\vb r} = \hat{\vb x}_1 - \hat{\vb x}_2$. The primitive electric quadrupole tensor becomes
\begin{equation}
    \hat q^{\mu\nu} = \frac{q}{4}\hat r^{\mu} \hat r^{\nu},
\end{equation}
which depends only on the relative coordinate. 
The expectation value of $\hat q^{\mu\nu}$ with $\ket{\Psi}$ is
\begin{equation}
    \bra{\Psi} \hat q^{\mu\nu} \ket{\Psi} = \frac{q}{4} \bra{\Psi} \hat r^{\mu} \hat r^{\nu}\ket{\Psi},
\end{equation}
Here, $\langle \hat r^{\mu}\hat r^{\nu}\rangle = \bra{\Psi} \hat r^{\mu} \hat r^{\nu}\ket{\Psi}$ is the Cooper pair quadrupole moment we compute, given by
\begin{equation}
\begin{aligned}
    \langle \hat r^{\mu}\hat r^{\nu}\rangle
    = \frac{1}{\mathcal{N}}\int d^2 R d^2r\sum_{\alpha\beta} |\psi_{\alpha\beta}(\vb{R},\vb r)|^2 r^\mu r^\nu,
\end{aligned}
\end{equation}
where the normalization is given in Eq.~\ref{normalization}.
%\JY{Wait, where does the center of mass coordinate go?}
%This shows that writing Haldane's definition in the two-body way and using the Cooper pair state $\ket{\Psi}$ to calculate it leads to the same Cooper pair quadrupole as we consider (up to a $q/4$ constant factor).
}  
The probability density can be written as
\begin{equation}
\begin{aligned}
    \sum_{\alpha\beta} |\psi_{\alpha\beta}|^2 = &\sum_{\vb{k}'\vb{k}} \psi(\vb{k}') \psi^*(\vb{k}) e^{i(\vb{k}'-\vb{k})\cdot \vb{r}} \sum_{\alpha} u_{\alpha\vb{x}_1}(\vb{k}') u^*_{\alpha\vb{x}_1}(\vb{k}) \sum_{\beta} u_{\beta\vb{x}_2}(-\vb{k}') u^*_{\beta\vb{x}_2}(-\vb{k})\\
    \approx &\sum_{\vb{k}'\vb{k}} \psi(\vb{k}') \psi^*(\vb{k}) e^{i(\vb{k}'-\vb{k})\cdot \vb{r}} \bra{u(\vb{k})}\ket{u(\vb{k}')} \bra{u(-\vb{k})}\ket{u(-\vb{k}')},
\end{aligned}
\end{equation}
In the tight-binding (localized-orbital) approximation we treat the cell-periodic parts as approximately position independent within a unit cell and use orthonormality of local orbitals, so that the real-space sums reduce to Bloch-state overlaps (form factors) $\sum_{\alpha}u_{\alpha}(\vb k')u_{\alpha}^*(\vb k)\simeq\langle u(\vb k)|u(\vb k')\rangle$, and similarly for $-\vb k$.
\WQC{The probability density thus depends only on the relative coordinate.} 
The Bloch overlap $\Lambda(\vb{k},\vb{k}') = \bra{u(\vb{k})}\ket{u(\vb{k}')}$ is also known as the form factor, which encodes the band geometry.

Then the Cooper pair quadrupole is given by
\begin{equation}
\begin{aligned}
    \langle r^\mu r^\nu \rangle = \frac{1}{\mathcal{N}}\int d^2 R d^2r\, (r^\mu r^\nu) \sum_{\vb{k}'\vb{k}} \psi(\vb{k}') \psi^*(\vb{k}) e^{i(\vb{k}'-\vb{k})\cdot \vb{r}} \Lambda_1(\vb{k},\vb{k}') \Lambda_2(\vb{k},\vb{k}').
\end{aligned}
\end{equation}
By defining $\vb{q} = \vb{k}' - \vb{k}$ and replacing $r^\mu r^\nu e^{i\vb{q}\cdot \vb{r}}\to -\partial_{q^\mu} \partial_{q^\nu} e^{i\vb{q}\cdot \vb{r}}$, we have
\begin{equation}
\begin{aligned}
    &\langle r^\mu r^\nu \rangle = \\
    &-\frac{1}{\mathcal{N}}\sum_{\vb{k},\vb{q}} \psi(\vb{k}+\frac{\vb{q}}{2}) \psi^*(\vb{k} - \frac{\vb{q}}{2}) [\Lambda_1(\vb{k} - \frac{\vb{q}}{2},\vb{k}+\frac{\vb{q}}{2}) \Lambda_2(\vb{k} - \frac{\vb{q}}{2},\vb{k}+\frac{\vb{q}}{2})] \partial_{q^\mu} \partial_{q^\nu} \delta(\vb{q}).
\end{aligned}
\end{equation}
Integrating by parts, shifting the derivatives, and setting $\vb{q} = 0$, we obtain
\begin{equation}
\begin{aligned}
   \langle r^\mu r^\nu \rangle =& -\frac{1}{\mathcal{N}}\sum_{\vb{k}} \partial_{q_\mu} \partial_{q_\nu} [\psi(\vb{k}+\frac{\vb{q}}{2}) \psi^*(\vb{k} - \frac{\vb{q}}{2})]|_{\vb{q}=0}\\
    &- \frac{1}{\mathcal{N}}\sum_{\vb{k}} |\psi(\vb{k})|^2 \partial_{q_\mu} \partial_{q_\nu} [\Lambda_1(\vb{k} - \frac{\vb{q}}{2},\vb{k}+\frac{\vb{q}}{2}) \Lambda_2(\vb{k} - \frac{\vb{q}}{2},\vb{k}+\frac{\vb{q}}{2})] |_{\vb{q}=0}\\
    &-\frac{1}{\mathcal{N}}\sum_{\vb{k}} \left[
    \partial_{q_\mu}[\psi(\vb{k}+\frac{\vb{q}}{2}) \psi^*(\vb{k} - \frac{\vb{q}}{2})] \partial_{q_\nu} [\Lambda_1(\vb{k} - \frac{\vb{q}}{2},\vb{k}+\frac{\vb{q}}{2}) \Lambda_2(\vb{k} - \frac{\vb{q}}{2},\vb{k}+\frac{\vb{q}}{2})]\right.\\
    &\left.\hspace{3.5cm}+\partial_{q_\nu}[\psi(\vb{k}+\frac{\vb{q}}{2}) \psi^*(\vb{k} - \frac{\vb{q}}{2})] \partial_{q_\mu} [\Lambda_1(\vb{k} - \frac{\vb{q}}{2},\vb{k}+\frac{\vb{q}}{2}) \Lambda_2(\vb{k} - \frac{\vb{q}}{2},\vb{k}+\frac{\vb{q}}{2})]\right]_{\vb{q}=0}.
\end{aligned}
\end{equation}
The first term can be written as
\begin{equation}
\begin{aligned}
    &-\frac{1}{\mathcal{N}}\sum_{\vb{k}} \partial_{q_\mu}\partial_{q_\nu} [\psi(\vb{k}+\frac{\vb{q}}{2}) \psi^*(\vb{k} - \frac{\vb{q}}{2})]|_{\vb{q}=0}\\
    =& -\frac{1}{\mathcal{N}}\sum_{\vb{k}} \frac{1}{4}[\psi^*\partial_{k_\mu}\partial_{k_\nu} \psi + \psi \partial_{k_\mu}\partial_{k_\nu} \psi^* - \partial_{k_\mu}\psi^*\partial_{k_\nu} \psi - \partial_{k_\nu}\psi^*\partial_{k_\mu} \psi]
\end{aligned}
\end{equation}
Assuming periodic boundary conditions in the Brillouin zone, or that the envelope function vanishes at the boundary, the second-derivative terms can be integrated by parts in $\vb{k}$. This gives
\begin{equation}
\begin{aligned}
    &-\frac{1}{\mathcal{N}}\sum_{\vb{k}} \partial_{q_\mu}\partial_{q_\nu} [\psi(\vb{k}+\frac{\vb{q}}{2}) \psi^*(\vb{k} - \frac{\vb{q}}{2})]|_{\vb{q}=0}\\
    =& \frac{1}{\mathcal{N}}\sum_{\vb{k}} \frac{1}{2}[\partial_{k_\mu}\psi^*\partial_{k_\nu} \psi + \partial_{k_\nu}\psi^*\partial_{k_\mu} \psi].
\end{aligned}
\end{equation}
The derivative term of $\psi(\vb{k})$ in the last term can be written as
\begin{equation}
\begin{aligned}
    \partial_{q_\mu}[\psi(\vb{k}+\frac{\vb{q}}{2}) \psi^*(\vb{k} - \frac{\vb{q}}{2})]|_{\vb{q}=0} = \frac{1}{2}[\partial_{k_\mu} \psi(\vb{k}) \psi^*(\vb{k}) - \psi(\vb{k}) \partial_{k_\mu} \psi^*(\vb{k})]
\end{aligned}
\end{equation}
To evaluate the derivatives of the form factors, we expand them to the second order in $\vb{q}$,
\begin{equation}
\begin{aligned}
    \Lambda_1(\vb{k} - \frac{\vb{q}}{2},\vb{k} + \frac{\vb{q}}{2}) = &\bra{u(\vb{k} - \frac{\vb{q}}{2})}\ket{u(\vb{k}+\frac{\vb{q}}{2})}\\
    =& 1 - \frac{q_{\mu}}{2} \bra{\partial_{k_\mu} u(\vb{k})}\ket{ u (\vb{k})} + \frac{q_\mu}{2} \bra{u(\vb{k})}\ket{ \partial_{k_\mu} u (\vb{k})}\\
     &+ \frac{q_\mu q_\nu}{8} \bra{\partial_{k_\mu} \partial_{k_\nu} u(\vb{k})}\ket{  u (\vb{k})} + \frac{q_\mu q_\nu}{8} \bra{ u(\vb{k})}\ket{\partial_{k_\mu} \partial_{k_\nu}  u (\vb{k})}\\
    &- \frac{q_\mu q_\nu}{8}\bra{ \partial_{k_\mu} u(\vb{k})}\ket{\partial_{k_\nu}  u (\vb{k})} - \frac{q_\mu q_\nu}{8}\bra{ \partial_{k_\nu} u(\vb{k})}\ket{\partial_{k_\mu}  u (\vb{k})}\\
    = &1 - i A^\mu(\vb{k}) q_\mu + \frac{q_\mu q_\nu}{4} \Re\bra{u(\vb{k})}\ket{\partial_{\mu}\partial_{\nu} u(\vb{k})} - \frac{q_\mu q_\nu}{4} \Re\bra{\partial_{\mu} u(\vb{k})}\ket{\partial_{\nu} u(\vb{k})} + \mathcal{O}(q^3),
\end{aligned}
\end{equation}
where $A^\mu(\vb{k}) = i \bra{u(\vb{k})}\ket{\partial_{k_\mu} u(\vb{k})}$ is the Berry connection.
Since $\bra{u(\vb{k})}\ket{\partial_{k_\mu} u(\vb{k})}$ is purely imaginary, we have $\Re\bra{ u(\vb{k})}\ket{\partial_{\mu} \partial_{\nu} u(\vb{k})} = -\Re\bra{\partial_{\mu} u(\vb{k})}\ket{\partial_{\nu} u(\vb{k})}$.
The form factor becomes
\begin{equation}
\begin{aligned}
    \Lambda_1(\vb{k} - \frac{\vb{q}}{2},\vb{k} + \frac{\vb{q}}{2}) = &\bra{u(\vb{k} - \frac{\vb{q}}{2})}\ket{u(\vb{k}+\frac{\vb{q}}{2})} \\
    = &1 - i A^\mu(\vb{k}) q_\mu - \frac{q_\mu q_\nu}{2} \Re\bra{\partial_{\mu} u(\vb{k})}\ket{\partial_{\nu} u(\vb{k})} + \mathcal{O}(q^3)\\
    = &1 - i A^\mu(\vb{k}) q_\mu - \frac{q_\mu q_\nu}{2} [g^{\mu\nu}(\vb{k}) + A^\mu(\vb{k}) A^\nu(\vb{k})] + \mathcal{O}(q^3)\\
    \approx &e^{-i A^\mu(\vb{k}) q_\mu - \frac{1}{2} g^{\mu\nu}(\vb{k})q_\mu q_\nu} 
\end{aligned}
\end{equation}
We note that the last equality is precise when Taylor expanding the exponential to the second order in $\vb{q}$: $e^{X}=1+X+\frac{1}{2}X^2+\mathcal O(q^3)$. 
The quantum metric $g^{\mu\nu}(\vb{k})$ is given by
\begin{equation}
\begin{aligned}
    g^{\mu\nu}(\vb{k}) = \Re\bra{\partial_{k_\mu}u(\vb{k})}\ket{\partial_{k_\nu}u(\vb{k})} - A^\mu(\vb{k}) A^\nu(\vb{k}).
\end{aligned}
\end{equation}
In a similar way, we have
\begin{equation}
\begin{aligned}
    \Lambda_2(\vb{k} - \frac{\vb{q}}{2},\vb{k} + \frac{\vb{q}}{2}) = &\bra{u(-\vb{k} + \frac{\vb{q}}{2})}\ket{u(-\vb{k} - \frac{\vb{q}}{2})}\\
    = &1 + i A^\mu(-\vb{k}) q_\mu - \frac{q_\mu q_\nu}{2} [g^{\mu\nu}(-\vb{k}) + A^\mu(-\vb{k}) A^\nu(-\vb{k})] + \mathcal{O}(q^3)\\
    \approx &e^{i A^\mu(-\vb{k}) q_\mu - \frac{1}{2} g^{\mu\nu}(-\vb{k})q_\mu q_\nu} 
\end{aligned}
\end{equation}
The pair form factor is given by the product of the two single-particle form factors,
\begin{equation}
\begin{aligned}
    \Lambda_{P} = \Lambda_1 \Lambda_2 = &1 - i \left[A^\mu(\vb{k}) - A^\mu(-\vb{k})\right] q_\mu + A^\mu(\vb{k}) A^\nu(-\vb{k}) q_\mu q_\nu\\
    &- \frac{1}{2} \left[g^{\mu\nu}(\vb{k}) + g^{\mu\nu}(-\vb{k}) + A^\mu(\vb{k}) A^\nu(\vb{k}) + A^\mu(-\vb{k}) A^\nu(-\vb{k})\right] q_\mu q_\nu + \mathcal{O}(q^3)
\end{aligned}
\end{equation}
The first and second order derivatives of the form factor product are
\begin{equation}
\begin{aligned}
    \partial_{q_\mu} \Lambda_{P} |_{\vb{q}=0} = - i [A^\mu(\vb{k}) - A^\mu(-\vb{k})]
\end{aligned}
\end{equation}
\begin{equation}
\begin{aligned}
    \partial_{q_\mu}\partial_{q_\nu} \Lambda_{P} |_{\vb{q}=0} = - [g^{\mu\nu}(\vb{k}) + g^{\mu\nu}(-\vb{k})] - [A^\mu(\vb{k}) - A^\mu(-\vb{k})][A^\nu(\vb{k}) - A^\nu(-\vb{k})]
\end{aligned}
\end{equation}
We can obtain the Cooper pair quadrupole as
\begin{equation}
\begin{aligned}
    \langle r^\mu r^\nu\rangle &= \frac{1}{\mathcal{N}}\sum_{\vb{k}} \frac{1}{2}[\partial_{k_\mu}\psi^*\partial_{k_\nu} \psi + \partial_{k_\nu}\psi^*\partial_{k_\mu} \psi] \\
    &+ \frac{1}{\mathcal{N}}\sum_{\vb{k}} \frac{i}{2} [\psi^* \partial_{k_\mu} \psi - \psi \partial_{k_\mu} \psi^*] [{A}^\nu(\vb{k}) - {A}^\nu(-\vb{k})]\\
    &+ \frac{1}{\mathcal{N}}\sum_{\vb{k}} \frac{i}{2} [\psi^* \partial_{k_\nu} \psi - \psi \partial_{k_\nu} \psi^*] [{A}^\mu(\vb{k}) - {A}^\mu(-\vb{k})]\\
    &+ \frac{1}{\mathcal{N}}\sum_{\vb{k}} |\psi(\vb{k})|^2 [A^\mu(\vb{k}) - A^\mu(-\vb{k})][A^\nu(\vb{k}) - A^\nu(-\vb{k})] \\
    &+ \frac{1}{\mathcal{N}}\sum_{\vb{k}}|\psi(\vb{k})|^2 [g^{\mu\nu}(\vb{k}) + g^{\mu\nu}(-\vb{k})]
\end{aligned}
\end{equation}
This expression is symmetric under $\mu\leftrightarrow\nu$ and can be written compactly as
\begin{equation}
\begin{aligned}
    \langle r^\mu r^\nu \rangle =\frac{1}{\mathcal{N}}\sum_{\vb{k}} \frac12\left[\mathcal{D}^*_\mu \psi^*(\vb{k}) \mathcal{D}_\nu \psi(\vb{k}) + \mathcal{D}^*_\nu \psi^*(\vb{k}) \mathcal{D}_\mu \psi(\vb{k}) \right] + \frac{1}{\mathcal{N}}\sum_{\vb{k}}|\psi(\vb{k})|^2 [g^{\mu\nu}(\vb{k}) + g^{\mu\nu}(-\vb{k})]
\end{aligned}
\end{equation}
\begin{equation}
\begin{aligned}
    \mathcal{D}_\mu = \partial_{k_\mu} - i[A^\mu(\vb{k}) - A^\mu(-\vb{k})]
\end{aligned}
\end{equation}
where $\vb{A}(\vb{k})$ and $\vb{A}(-\vb{k})$ are the Berry connections of the two electrons at $\vb{k}$ and $-\vb{k}$, respectively, and $g^{\mu\nu}(\vb{k})$ and $g^{\mu\nu}(-\vb{k})$ are the quantum metrics of the two electrons in the Cooper pair.
We define the pair Berry connection as
\begin{equation}
\begin{aligned}
    \vb{A}_{P}(\vb{k}) = \vb{A}(\vb{k}) - \vb{A}(-\vb{k})
\end{aligned}
\end{equation}
and the pair quantum metric as
\begin{equation}
\begin{aligned}
    g_{P}^{\mu\nu}(\vb{k}) = g^{\mu\nu}(\vb{k}) + g^{\mu\nu}(-\vb{k})
\end{aligned}
\end{equation}

The Cooper pair envelope function $\psi(\vb{k})$ can be written as $\psi(\vb{k}) = |\psi(\vb{k})| e^{i\phi(\vb{k})}$, where $\phi(\vb{k})$ is the pair phase.
Under the gauge transformation of the Bloch states $\ket{u(\vb{k})} \to e^{i\alpha(\vb{k})}\ket{u(\vb{k})}$, the band creation operators transform as
\begin{equation}
\begin{aligned}
    a^\dagger_{\vb{k}}  \to e^{i\alpha(\vb{k})} a^\dagger_{\vb{k}}
\end{aligned}
\end{equation}
Through the Cooper pair creation operator $\mathcal{P}^\dagger = \sum_{\vb{k}} \psi(\vb{k}) a^\dagger_{\vb{k}\uparrow} a^\dagger_{-\vb{k}\downarrow}$, the Cooper pair envelope function $\psi(\vb{k})$ must be gauge-covariant, transforming as
\begin{equation}
\begin{aligned}
    \psi(\vb{k}) \to \psi(\vb{k})  e^{-i(\alpha_+(\vb{k}) + \alpha_-(-\vb{k}))}
\end{aligned}
\end{equation}
so that $\mathcal{P}^\dagger$ is invariant.  Here $\alpha_+(\vb{k})$ and $\alpha_-(-\vb{k})$ are the gauge phases for the two electrons in the Cooper pair at $\vb{k}$ and $-\vb{k}$, respectively.
Therefore, the phase of the Cooper pair envelope function transforms as
\begin{equation}
\begin{aligned}
    \phi(\vb{k}) \to \phi(\vb{k}) - \alpha_+(\vb{k}) - \alpha_-(-\vb{k})
\end{aligned}
\end{equation}
The pair Berry connection transforms as
\begin{equation}
\begin{aligned}
    \vb{A}_{P}(\vb k) \to \vb{A}_{P}(\vb{k}) - \nabla_{\vb{k}} [\alpha_+(\vb{k}) + \alpha_-(-\vb{k})]
\end{aligned}
\end{equation}
From the gauge transformations, we can see that the combination 
\begin{equation}
\begin{aligned}
    \vb{d}_\text{G}(\vb k) = \nabla_{\vb{k}} \phi(\vb{k}) - \vb{A}_{P}(\vb{k})
\end{aligned}
\end{equation}
is gauge-invariant \cite{Topological-Nodal-Cooper}.  
It is the k-space analogue of the gauge-invariant superfluid velocity in real space $\vb{v}_s = \nabla_{\vb{r}} \phi_{\vb r} - \frac{2e}{\hbar}\bm{A}(\vb r)$. $\vb{d}_\text{G}(\vb k)$ can be viewed as the pair geometric dipole.

Then we can define a gauge-invariant \textbf{quantum geometric quadrupole} moment tensor of the Cooper pair as
\begin{equation}
\begin{aligned}
    \mathcal{Q}^{\mu\nu}_\text{G}(\vb{k}) = [\partial_{k_\mu} \phi(\vb{k}) - A^\mu_{P}(\vb{k})][\partial_{k_\nu} \phi(\vb{k}) - A^\nu_{P}(\vb{k})] + g^{\mu\nu}_{P}(\vb{k})
\end{aligned}
\end{equation}
\begin{equation}
\begin{aligned}
    \langle r^\mu r^\nu \rangle_G &= \frac{1}{\mathcal{N}}\sum_{\vb{k}} |\psi(\vb{k})|^2 \mathcal{Q}^{\mu\nu}_{G}(\vb{k})\\
    & = \frac{1}{\mathcal{N}}\sum_{\vb{k}} |\psi(\vb{k})|^2 \left\{[\partial_{k_\mu} \phi(\vb{k}) - A^\mu_{P}(\vb{k})][\partial_{k_\nu} \phi(\vb{k}) - A^\nu_{P}(\vb{k})] + g^{\mu\nu}_{P}(\vb{k})\right\}
\end{aligned}
\end{equation}
Together with the contribution from the amplitude variation of $\psi(\vb{k})$, 
\begin{equation}
\begin{aligned}
    \langle r^\mu r^\nu \rangle_\text{A} = \frac{1}{\mathcal{N}}\sum_{\vb{k}} \frac{1}{2}\left[\partial_{k_\mu} |\psi(\vb{k})| \partial_{k_\nu} |\psi(\vb{k})| - |\psi(\vb{k})| \partial_{k_\mu} \partial_{k_\nu} |\psi(\vb{k})|\right]
\end{aligned}
\end{equation}
we have the total pair quadrupole $\langle r^\mu r^\nu\rangle = \langle r^\mu r^\nu \rangle_G + \langle r^\mu r^\nu \rangle_A$ expressed in terms of the gauge-invariant quantities.

The trace of the quadrupole is the Cooper pair size,
\begin{equation}
\begin{aligned}
    \xi_{\text{Pair}}^2 = \xi^2_{\text{Amp}} + \xi^2_{\text{G}}
\end{aligned}
\end{equation}
\begin{equation}
\begin{aligned}
    \xi_{\text{Amp}}^2
    = \frac{1}{\mathcal{N}}\sum_{\vb{k}}\frac{1}{2}
    \left[
    \qty(\nabla_{\vb{k}} |\psi(\vb{k})|)^2
    - |\psi(\vb{k})| \nabla_{\vb{k}}^2 |\psi(\vb{k})|
    \right].
\end{aligned}
\end{equation}
When boundary terms vanish, this expression is equivalent to
$\mathcal{N}^{-1}\sum_{\vb{k}}\qty(\nabla_{\vb{k}}|\psi(\vb{k})|)^2$.
\begin{equation}
\begin{aligned}
    \xi_{\text{G}}^2 = \xi^2_{\text{Phase}} + \xi^2_{\text{Metric}} = \frac{1}{\mathcal{N}}\sum_{\vb{k}} |\psi(\vb{k})|^2 \left[\nabla_{\vb{k}} \phi(\vb{k}) - \vb{A}_{P}(\vb{k})\right]^2 + \frac{1}{\mathcal{N}}\sum_{\vb{k}} |\psi(\vb{k})|^2 \Tr[g^{\mu\nu}_{P}(\vb{k})]
\end{aligned}
\end{equation}
This expression clarifies how to compute the Cooper pair size from the pair envelope function $\psi(\vb{k})$ and the quantum geometry of the band.

\section{Geometric bound of Cooper pair size}

In a Brillouin zone with periodic boundary conditions, the amplitude contribution to the pair size can be integrated by parts, and the boundary terms vanish. The projected squared relative-coordinate operator then takes the form
\begin{equation}
\begin{aligned}
\hat r^2
= \sum_{\vb k} \ket{\vb k_P}\Bigl[
\mathcal{D}_{\vb k}^\dagger\cdot \mathcal{D}_{\vb k}
+
\Tr g^{\mu\nu}_P(\vb k)\Bigr] \bra{\vb k_P},
\end{aligned}
\end{equation}
where $\Tr g^{\mu\nu}_P\equiv\delta_{\mu\nu}g_P^{\mu\nu}$ and
\(
\mathcal{D}_{\vb k}=\nabla_{\vb k}-i\vb A_P(\vb k)
\)
is the pair covariant derivative. Defining the Hermitian operator
\(
D_{\vb k}=i\mathcal{D}_{\vb k}
= i\nabla_{\vb k}+\vb A_P(\vb k),
\)
we have
\begin{equation}
\begin{aligned}
\hat r^2
= \sum_{\vb k} \ket{\vb k_P}\Bigl[
D_{\vb k}^2+\Tr g_P^{\mu\nu}(\vb k)\Bigr] \bra{\vb k_P}.
\end{aligned}
\end{equation}

This operator is formally analogous to the Hamiltonian of a charged particle moving in a magnetic field in real space. In this analogy, the pair Berry curvature
\(
\Omega_P(\vb k)=\nabla_{\vb k}\!\times\!\vb A_P(\vb k)
\)
acts as an effective magnetic field in momentum space, while the trace of the pair quantum metric, $\Tr g_P^{\mu\nu}$, plays the role of a scalar potential. The commutator of the covariant derivative components is
\begin{equation}
\begin{aligned}
\big[D_{k_x},D_{k_y}\big]
=
i\,\Omega_P(\vb k).
\end{aligned}
\end{equation}
Introducing holomorphic and antiholomorphic operators,
\begin{equation}
\begin{aligned}
D=\frac{D_{k_x}+iD_{k_y}}{2},\qquad
\bar D=\frac{D_{k_x}-iD_{k_y}}{2},
\end{aligned}
\end{equation}
we obtain
\begin{equation}
\begin{aligned}
[\,D,\bar D\,]
=
\frac{1}{2}\,\Omega_P(\vb k),
\end{aligned}
\end{equation}
which allows us to complete the square in two equivalent ways:
\begin{equation}
\begin{aligned}
D_{\vb k}^2
=
2(D\bar D+\bar D D)
=
4\bar D D+\Omega_P(\vb k)
=
4 D\bar D - \Omega_P(\vb k).
\end{aligned}
\end{equation}
Now consider a normalized Cooper pair state whose envelope function is supported, or strongly concentrated, in a momentum-space region $\mathcal R$.
We assume that the pair Berry curvature does not change sign in this region. For definiteness, take $\Omega_P(\vb k)\ge0$ in $\mathcal R$.
Using the ordering
\(
D_{\vb k}^2=4\bar D D+\Omega_P(\vb k)
\),
the expectation value of $\hat r^2$ becomes
\begin{equation}
\begin{aligned}
\langle \hat r^2\rangle
&=
\sum_{\vb k\in\mathcal R}
    \psi_{\vb k}^*
    \Bigl[4\bar D D+\Omega_P(\vb k)+\Tr g^{\mu\nu}_P(\vb k)\Bigr]
    \psi_{\vb k}\\
&=
4\langle D\Psi|D\Psi\rangle
+
\sum_{\vb k\in\mathcal R}
|\psi_{\vb k}|^2
\Bigl[\Omega_P(\vb k)+\Tr g_P(\vb k)\Bigr].
\end{aligned}
\end{equation}
Since the first term is nonnegative, the Cooper pair size obeys
\begin{equation}
\begin{aligned}
\langle \hat r^2\rangle
\ge
\sum_{\vb k\in\mathcal R} |\psi_{\vb k}|^2
\Bigl[\Omega_P(\vb k)+\Tr g_P(\vb k)\Bigr]
\equiv \xi^2_{\rm Bound}.
\end{aligned}
\end{equation}
If instead $\Omega_P(\vb k)\le0$ throughout $\mathcal R$, one uses the equivalent ordering
\(
D_{\vb k}^2=4D\bar D-\Omega_P(\vb k)
\)
and obtains the corresponding result with $\Omega_P$ replaced by $-\Omega_P$. Below we keep the convention $\Omega_P(\vb k)\ge0$ in $\mathcal R$.
Since the envelope is normalized within the region $\mathcal R$, the right-hand side is bounded below by the minimum value of the local geometric contribution in that region,
\begin{equation}
\begin{aligned}
\langle \hat r^2\rangle
\ge
\min_{\vb k\in\mathcal R}
\Bigl\{
\Omega_P(\vb k)+\Tr g_P(\vb k)
\Bigr\}
\equiv
\xi_{\rm fund}^2.
\end{aligned}
\end{equation}
This $\xi_{\rm fund}$ is a geometry-only lower bound associated with the sign-definite region $\mathcal R$. It is weaker than the envelope-weighted bound above, but it is independent of the detailed pairing channel as long as the Cooper pair envelope is supported, or strongly concentrated, in $\mathcal R$.

The bound can be further simplified using the trace inequality. The single-particle trace inequality
\(
\Tr g(\vb k)\ge|\Omega(\vb k)|
\)
implies the corresponding inequality for the pair geometric quantities,
\begin{equation}
\begin{aligned}
    \Tr g^{\mu\nu}_{P}(\vb{k}) = \Tr g^{\mu\nu}(\vb{k}) + \Tr g^{\mu\nu}(-\vb{k})
    \ge |\Omega(\vb{k})| + |\Omega(-\vb{k})|
    \ge |\Omega_{P}(\vb{k})|.
\end{aligned}
\end{equation}
In the chosen sign convention, $\Omega_P(\vb k)\ge0$ in $\mathcal R$, this becomes $\Tr g_P(\vb k)\ge \Omega_P(\vb k)$ in $\mathcal R$. Applying it to the envelope-weighted bound gives
\begin{equation}
\begin{aligned}
\langle \hat r^2\rangle
\ge
2\sum_{\vb k\in\mathcal R} |\psi_{\vb k}|^2 \Omega_P(\vb k).
\end{aligned}
\end{equation}

\section{Two-band model}
In this section, we present the details of the Cooper pair size calculation for the two-band model of rhombohedral pentalayer graphene. The single-particle Hamiltonian in valley $K$ is
\begin{equation}
h(\mathbf{k})=
\begin{pmatrix}
\frac{\Delta}{2}+\gamma_{-}\!\left(k_x^{2}+k_y^{2}\right)
&
\dfrac{v_F^{5}}{t_1^{4}}\,(k_x-\mathrm{i}k_y)^{5}
\\[6pt]
\dfrac{v_F^{5}}{t_1^{4}}\,(k_x+\mathrm{i}k_y)^{5}
&
-\frac{\Delta}{2}+\gamma_{+}\!\left(k_x^{2}+k_y^{2}\right)
\end{pmatrix},
\end{equation}
where $v_F = \frac{\sqrt{3}}{2} t_0 a$, $a = 0.246$ nm, $t_{0} = 3.1 \, \text{eV}$,
$t_{1} = 380 \, \text{meV}$, 
$t_{2} = 141 \, \text{meV}$,
$\gamma_{\pm} \approx \frac{2 v_F v_2}{t_1} \pm \frac{\Delta v_F^{2}}{4 t_1^{2}}$, and $v_2 = \frac{\sqrt{3}}{2}\, t_2 a$.
The corresponding band dispersions are
\begin{equation}
E_{\pm}(\mathbf{k})
= \frac{\gamma_{+}+\gamma_{-}}{2}\,k^{2}
\;\pm\;
\sqrt{\Bigl(\frac{\Delta}{2}+\frac{\gamma_{-}-\gamma_{+}}{2}\,k^{2}\Bigr)^{2}
+ \left(\frac{v_F^{5}}{t_1^{4}}\right)^{2} k^{10}},
\qquad k=\sqrt{k_x^2+k_y^2}.
\end{equation}
The conduction-band eigenvector is
\begin{equation}
\ket{u(\mathbf{k})}
=\frac{1}{\sqrt{2|\mathbf{d}|\bigl(|\mathbf{d}|+d_z\bigr)}}
\begin{pmatrix}
|\mathbf{d}|+d_z \\[6pt]
\dfrac{v_F^{5}}{t_1^{4}}\, (k_x+i k_y)^{5}
\end{pmatrix},
\end{equation}
where
\begin{equation}
|\mathbf{d}|=\sqrt{d_z^2+\left(\frac{v_F^{5}}{t_1^{4}}\,k^{5}\right)^2}, 
\quad d_z = \frac{\Delta}{2} + \frac{1}{2}(\gamma_{-}-\gamma_{+})k^2,
\quad k=\sqrt{k_x^2+k_y^2}.
\end{equation}

We restrict attention to intravalley, spin-polarized electronic states and consider a local attractive interaction projected onto the conduction-band basis. The projected interaction matrix element is
\begin{equation}
    V_{\vb{k}\vb{k}'} = - U \bra{u(\vb{k})}\ket{u(\vb{k}')} \bra{u(-\vb{k})}\ket{u(-\vb{k}')}.
\end{equation}
Because the pairing occurs within the same spin and valley sector, the interaction must be antisymmetrized \cite{Family-of-multilayer,From-Fractionalization-to-Chiral}, which gives
\begin{equation}
    \begin{aligned}
        V_{\vb{k}\vb{k}'} &= - \frac{U}{2} [\bra{u(\vb{k})}\ket{u(\vb{k}')} \bra{u(-\vb{k})}\ket{u(-\vb{k}')} - \bra{u(-\vb{k})}\ket{u(\vb{k}')} \bra{u(\vb{k})}\ket{u(-\vb{k}')} ]\\
    &= - U\frac{\Lambda^{P}(\vb{k},\vb{k}') - \Lambda^{P}(-\vb{k},\vb{k}')}{2}.
    \end{aligned}
\end{equation}
Using the explicit conduction-band spinor, we obtain
\begin{equation}
\begin{aligned}
    \Lambda^{P}(\vb{k},\vb{k}') = \frac{[(d + d_z)(d' + d_z') + \frac{v_F^{10}}{t_1^8}(k_x - i k_y)^5(k_x' + ik_y')^5]^2}{4 d d'(d + d_z)(d' + d_z')}, 
\end{aligned}
\end{equation}
\begin{equation}
\begin{aligned}
    \Lambda^{P}(-\vb{k},\vb{k}') = \frac{[(d + d_z)(d' + d_z') - \frac{v_F^{10}}{t_1^8}(k_x - i k_y)^5(k_x' + ik_y')^5]^2}{4 d d'(d + d_z)(d' + d_z')}, 
\end{aligned}
\end{equation}
\begin{equation}
\begin{aligned}
    \Lambda^{P}(\vb{k},\vb{k}') - \Lambda^{P}(-\vb{k},\vb{k}') = \frac{\frac{v_F^{10}}{t_1^8}(k_x - i k_y)^5(k_x' + ik_y')^5}{d d'},
\end{aligned}
\end{equation}
so that the antisymmetrized band-projected interaction becomes
\begin{equation}
\begin{aligned}
    V_{\vb{k}\vb{k}'} = -U \frac{\frac{v_F^{10}}{t_1^8}(k_x - i k_y)^5(k_x' + ik_y')^5}{2d d'},
\end{aligned}
\end{equation}
where $d = |\vb{d}(\vb{k})|$, $d' = |\vb{d}(\vb{k}')|$, $d_z = d_z(\vb{k})$, and $d_z' = d_z(\vb{k}')$.

\begin{figure*}
\includegraphics[width=0.35\linewidth]{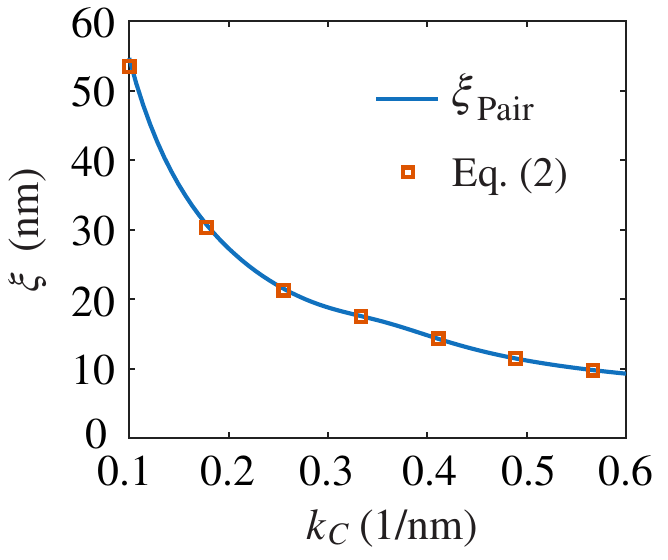}
    \caption{Comparison between the Cooper pair size obtained from the decomposition and the direct evaluation using Eq.~(2) of the main text.}\label{fig: compare}
\end{figure*}

Substituting the interaction and the Cooper pair state $\ket{\Psi} = \sum_{\vb{k}} \psi(\vb{k}) a^\dagger_{\vb{k}} a^\dagger_{-\vb{k}} \ket{GS}$ into the two-body Schr\"odinger equation $(H_0 + V)\ket{\Psi} = E_B \ket{\Psi}$, we obtain the pair eigenvalue equation
\begin{equation}
\begin{aligned}
    [ E_B - (2\varepsilon_{\vb{k}} - 2\varepsilon_F) ]\psi(\vb{k}) = \frac{1}{2S}\sum_{|\vb{k}'| < k_C} V_{\vb{k}\vb{k}'}\psi(\vb{k}').
\end{aligned}
\end{equation}
Here $S$ is the two-dimensional sample area. In the flat-band limit, we set $\varepsilon_{\vb{k}} - \varepsilon_F = 0$, so the equation reduces to
\begin{equation}
\begin{aligned}
    E_B \psi(\vb{k}) = \frac{1}{2S}\sum_{|\vb{k}'| < k_C} V_{\vb{k}\vb{k}'}\psi(\vb{k}'),
\end{aligned}
\end{equation}
\begin{equation}
\begin{aligned}
    E_B \psi(\vb{k}) = -U \frac{v_F^{10}}{2t_1^8}\frac{(k_x - i k_y)^5}{d}\frac{1}{2S}\sum_{|\vb{k}'| < k_C} \frac{(k_x' + i k_y')^5}{d'} \psi(\vb{k}'). 
\end{aligned}
\end{equation}
Because $V_{\vb k,\vb k'}$ factorizes into independent functions of \(\vb k\) and \(\vb k'\), the pair envelope function can be obtained analytically as
\begin{equation}
\begin{aligned}
    \psi(\vb{k}) = \frac{1}{\sqrt{\mathcal{N}}}\frac{ (k_x - i k_y)^5}{|\vb{d}(\vb{k})|}, 
\end{aligned}
\end{equation}
where $\mathcal{N}$ is the normalization factor.

To verify the decomposition of the Cooper pair size, we compare it with the direct evaluation of $\sqrt{\langle r^2\rangle}$ from Eq.~(2) of the main text. As shown in Fig.~S\ref{fig: compare}, the two results agree for all $k_C$, confirming that $\xi_{\rm Pair}=\sqrt{\xi_{\rm Phase}^2+\xi_{\rm Amp}^2+\xi_{\rm Metric}^2}$ is exact.

\section{Berry trashcan model}

\begin{figure*}
\includegraphics[width=0.8\linewidth]{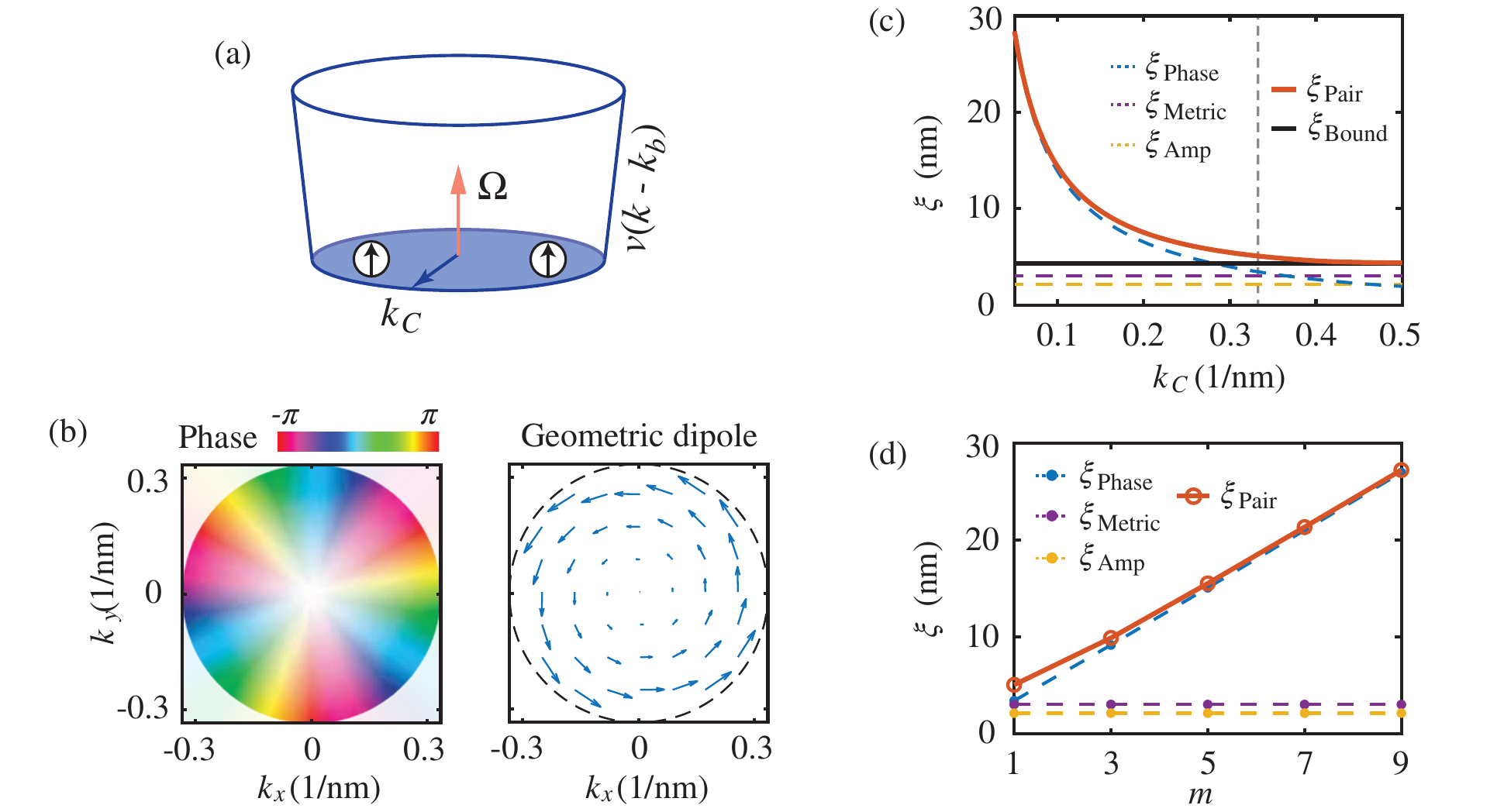}
    \caption{Cooper pair size in the Berry trashcan model for rhombohedral graphene. (a) Schematic low-energy dispersion with a flat-band bottom of radius $k_C$, enclosing a uniform Berry curvature $\Omega$, and a dispersive wall with velocity $v$. (b) Phase of the pair envelope function $\psi_{m}(\vb k)$ for $m = 3$; pair geometric dipole $\vb d_G$ circulating around the vortex core. (c) Cooper pair size and its individual contributions as functions of the momentum cutoff, together with the geometric lower bound. Here $m = 1$, and the dashed vertical line marks $k_C = 0.334\,\rm nm^{-1}$ for pentalayer graphene. (d) Cooper pair size for different angular-momentum channels $m$ at $k_C = 0.334\,\rm nm^{-1}$.}\label{fig:berry trashcan}
\end{figure*}

In this section, we present the details of the Cooper pair size calculation for the Berry trashcan effective model for rhombohedral graphene introduced in Refs.~\cite{bernevig2025berrytrashcanmodelinteracting,li2025berrytrashcanshortrange}. This model captures the low-energy conduction band shown in Fig.~\ref{fig:berry trashcan}(a). Near valley $K$, the Hamiltonian is $\hat H = \hat h + \hat V$, with kinetic term
\begin{equation}
        \hat h = \sum_{\vb k} \varepsilon_{\vb k}\,
\hat a^\dagger_{\vb k} \hat a_{\vb k},
\quad
\varepsilon_{\vb k} = \Theta(k-k_C)\, v (k-k_C),
\end{equation}
where $k_C$ is the radius of the flat-band region, determined by the number of layers \cite{bernevig2025berrytrashcanmodelinteracting}, $v$ sets the slope of the dispersive part of the band, and $\Theta(x)$ is the step function.
The projected attractive interaction is
\begin{equation}
    \hat V = -\frac{U}{2S}\sum_{\vb{k},\vb{k}',\vb{q}}
e^{\alpha q^2}
\Lambda_{\vb{k},\vb{q}}
\Lambda^*_{\vb{k}',\vb{q}}
\hat a^\dagger_{\vb{k}+\vb{q}}
\hat a^\dagger_{\vb{k}'-\vb{q}}
\hat a_{\vb{k}'} \hat a_{\vb{k}},
\end{equation}
where $S$ is the system area and the form factor is
\begin{equation}
\Lambda_{\vb{k},\vb{q}}
= \exp[-i\beta(\vb q\times \vb k)
-\tfrac{1}{2}|\beta|q^2].
\end{equation}
The corresponding Berry curvature and quantum metric are $\Omega_{\vb k} = 2\beta$ and $g^{\mu\nu}_{\vb k} = |\beta|\,\delta_{\mu\nu}$, respectively, with $\beta = v_F^2 / t_1^2$ \cite{bernevig2025berrytrashcanmodelinteracting}.

Reference \cite{li2025berrytrashcanshortrange} studied intravalley spin-polarized pairing in this model and, in the limit $v = \infty$ and $\alpha = 0$, obtained the analytic pair envelope function
\begin{equation}
    \psi_m (\vb k) = \frac{1}{\sqrt{\mathcal{N}}}(k_x + i k_y)^m e^{-|\beta| k^2},
\end{equation}
where $m$ is an odd angular-momentum channel and $\mathcal{N}$ is the normalization factor.
Figure~S\ref{fig:berry trashcan}(b) shows the vortex structure of $\psi_m(\vb k)$, whose phase winds by $m$ around the origin.
We adopt the symmetric gauge for the Berry connection, $\vb A(\vb k) = \beta\,(-k_y,\,k_x)$. The corresponding pair Berry connection is
\begin{equation}
\vb A_P(\vb k)
    = \vb A(\vb k) - \vb A(-\vb k)
    = 2\beta\,(-k_y,\,k_x),
\end{equation}
and the pair Berry curvature is $\Omega_P = 2\Omega = 4\beta$.
As shown in Fig.~S\ref{fig:berry trashcan}(b), the pair geometric dipole $\vb d_G(\vb k) = \nabla_{\vb k}\phi(\vb k) - \vb A_P(\vb k)$ forms a circulating texture around the vortex core, with circulation
\begin{equation}
\oint d\vb k\cdot \vb d_G(\vb k)
= 2\pi m - 4\pi\beta k_C^2
= 2\pi m - \Phi_P,
\end{equation}
where $\Phi_P=\Omega_P\pi k_C^2=4\pi\beta k_C^2$ is the pair Berry flux enclosed by the flat-band region.

In Fig.~S\ref{fig:berry trashcan}(c), we treat the flat-band radius $k_C$ as a tunable model parameter and plot the Cooper pair size and its individual contributions as functions of $k_C$ for the $p$-wave channel ($m=1$).
The amplitude contribution is independent of both $k_C$ and $m$:
\begin{equation}
\begin{aligned}
    &|\psi(\vb k)| = k^{m} e^{-|\beta|k^2},\quad \partial_k |\psi(\vb k)|
    = k^{m-1} e^{-|\beta|k^2}\bigl(m-2|\beta|k^2\bigr),\\
    &\nabla_{\vb k}^2 |\psi(\vb k)|
    = k^{m-2} e^{-|\beta|k^2}
       \bigl(4|\beta|^2k^4 - 4|\beta|(m+1)k^2 + m^2\bigr),
\end{aligned}
\end{equation}
so that
\begin{equation}
\begin{aligned}
    \xi^2_{\rm Amp}
    &= \frac{1}{\mathcal N}
    \sum_{|\vb k|<k_C}
    \frac{1}{2}
    \Bigl[
        (\nabla_{\vb k}|\psi(\vb k)|)^2
        - |\psi(\vb k)|\,\nabla_{\vb k}^2|\psi(\vb k)|
    \Bigr]\\
    & = \frac{1}{\mathcal N} \sum_{|\vb k|<k_C} 2|\beta|\,k^{2m} e^{-2|\beta|k^2}=\frac{1}{\mathcal N} \sum_{|\vb k|<k_C} 2|\beta|\,|\psi(\vb k)|^2 = 2|\beta|.
\end{aligned}
\end{equation}
The metric contribution is likewise independent of $k_C$ and $m$. The pair quantum metric is
\begin{equation}
g^{\mu\nu}_P(\vb k)
    = g^{\mu\nu}(\vb k) + g^{\mu\nu}(-\vb k)
    = 2|\beta|\,\delta_{\mu\nu}.
\end{equation}
In two dimensions, this implies
\begin{equation}
\mathrm{Tr}\,g_P(\vb k)
    = \delta_{\mu\nu} g^{\mu\nu}_P(\vb k)
    = 4|\beta|,
\end{equation}
and therefore
\begin{equation}
\begin{aligned}
    \xi^2_{\rm Metric}
    = \frac{1}{\mathcal N}
      \sum_{|\vb k|<k_C}
      |\psi(\vb k)|^2\,
      \mathrm{Tr}\,g^{\mu\nu}_P(\vb k) = 4|\beta|.
\end{aligned}
\end{equation}
Thus, $\xi_{\rm Metric} = \sqrt{4|\beta|}$ and $\xi_{\rm Amp} = \sqrt{2|\beta|}$ remain constant, whereas the phase contribution $\xi_{\rm Phase}$ depends on $k_C$ and dominates when $k_C$ is small. The geometric lower bound $\xi_{\rm Bound} = \sqrt{8|\beta|}$ is also shown in Fig.~S\ref{fig:berry trashcan}(c). For the flat-band region, where the quantum geometry is constant and the envelope is supported in the sign-definite region, the envelope-weighted bound and the minimum bound coincide:
\begin{equation}
\begin{aligned}
    \xi_{\rm Bound}^2
    = \frac{1}{\mathcal N}
      \sum_{|\vb k|<k_C}
      |\psi(\vb k)|^2
      \bigl[
        |\Omega_P(\vb k)|
        + \mathrm{Tr}\,g_P(\vb k)
      \bigr]
    = \xi_{\rm fund}^2
    = 4|\beta| + 4|\beta|
    = 8|\beta|.
\end{aligned}
\end{equation}
Using $\beta\approx 2.33~\text{nm}^2$ gives
\begin{equation}
\xi_{\rm Bound}=\xi_{\rm fund}
    = \sqrt{8|\beta|}
    \approx 4.3~\text{nm}.
\end{equation}
In Fig.~S\ref{fig:berry trashcan}(d), we take $k_C = 0.334\,\rm nm^{-1}$ (the flat-band radius extracted for pentalayer~\cite{bernevig2025berrytrashcanmodelinteracting}) and compare the pair size for different angular momentum channels.
Higher winding number $m$ strongly enhances $\xi_{\rm Phase}$, demonstrating that increased chirality directly enlarges the Cooper pair size.

\end{document}